\documentclass[10pt]{article}
\usepackage{psfrag}
\usepackage[dvips]{graphicx}
\usepackage{euler}

\twocolumn
\begin{document}
\title{Thermodynamical fluctuations in optical mirror coatings}
\author{V. B. Braginsky and S. P. Vyatchanin\\ 	
    Physics Faculty, Moscow State University,
    Moscow 119899, Russia \\
    e-mail: vyat@hbar.phys.msu.su}
\date{\today}
\maketitle
\begin{abstract}
Thermodynamical (TD) fluctuations of temperature in mirrors may produce 
surface fluctuations not only through thermal
expansion in mirror body \cite{bgv} but also through thermal expansion in 
mirror coating. We analyze the last "surface" 
effect which can be larger than the first "volume" one due to larger thermal 
expansion coefficient of coating material and
smaller effective volume. In particular, these fluctuations may be  
important in laser interferometric
gravitational antennae.
\end{abstract}

\section{Introduction}
Outstanding experimental achievements in quantum optics and high 
resolution spectroscopy within recent four decades are
substantially due to creation and development of new technologies. 
The coating of mirrors, i.e. the deposition of many thin
dielectrical layers on the mirror surface is likely to be one of the 
most important technology in this part of experimental
physics. Today this technology provides the reflectivity $R$ of such 
coating very close to unity (at the beginning of laser era the value of
$(1-R)$ was about several percent while at present $(1-R)\simeq 10^{-6}$ 
\cite{rempe,kimble}). There is reason to hope that in the not too distant 
future
the value of $(1-R)$ will be close to $10^{-9}$. This improvement in particular means that measurement using squeezed quantum states will
become routine as the squeezed state lifetime is longer if losses are smaller. 
The rise of squeezing factor provides the
possibility to increase sensitivity with the same number of "used" photons 
\cite{bh1,bh2}.

The value of $(1-R)$ is of great concern in terrestrial gravitational wave 
antennae (projects LIGO and VIRGO, e.g. see \cite{abr1,abr2,bru}). These
antennae can be regarded as extremely sensitive spectrometers which goal is 
to detect very small deviations $\Delta\omega_{\rm FP}/\omega_{\rm FP}$
of one Fabry-Perot optical resonator mode eigen frequency in comparison with 
another one. The amplitude of these frequency relative deviations
which differs from the metric perturbations amplitude only by the factor of 
two, is quite small. The planned resolution for the stage LIGO-I
is $\Delta\omega_{\rm FP}/\omega_{\rm FP}\simeq10^{-21}$, and for the stage 
LIGO-II is one order smaller.

The above considerations show that the value of $(1-R)$ decrease is important 
for further development of many optical measuring devices
types. However, to our knowledge there is no deep analysis of possible 
additional noises which can appear specifically due to coating. In
this article we present the analysis of a noise source in the coated mirror, 
which deserves serious attention. Thermodynamical (TD)
fluctuations of temperature in mirror coating can produce additional 
fluctuations of surface through thermal expansion in mirror coating.
This noise may be larger than analogous one in mirror body due to larger 
coating materials thermal expansion coefficient and smaller
effective volume.
\begin{figure}
\unitlength0.1mm
\linethickness{0.5pt}
\begin{picture}(1423,682)
\put(356,479){\dashbox{20}(100,100)[]{}}
\put(396,459){\vector(-1,0){40}}
\put(396,459){\vector(1,0){60}}
\put(336,659){\line(1,0){440}}
\put(776,659){\line(0,-1){640}}
\put(776,19){\line(-1,0){440}}
\put(276,339){\vector(0,1){280}}
\put(276,339){\vector(0,-1){260}}
\put(216,379){$r_0$}
\bezier{560}(340,620)(620,620)(620,340)
\bezier{520}(620,340)(620,100)(340,100)
\put(560,580){\vector(-4,-1){160}}
\put(516,599){$\Delta T_{\rm TD}^2=\frac{k_B T^2}{\rho C r_T^3}$}
\bezier{169}(120,480)(40,440)(40,360)
\bezier{160}(40,360)(40,300)(120,240)
\bezier{173}(120,240)(180,200)(200,100)
\put(40,360){\vector(1,0){160}}
\put(45,400){\vector(1,0){155}}
\put(68,440){\vector(1,0){132}}
\put(123,480){\vector(1,0){77}}
\put(46,320){\vector(1,0){154}}
\put(73,280){\vector(1,0){127}}
\put(123,240){\vector(1,0){77}}
\put(175,201){\vector(1,0){25}}
\put(181,160){\vector(1,0){19}}
\bezier{189}(120,480)(200,520)(200,620)
\put(171,520){\vector(1,0){29}}
\put(191,560){\vector(1,0){9}}
\put(0,0){\framebox(860,680)[]{}}
\put(380,660){\line(0,-1){640}}
\bezier{222}(340,660)(320,560)(340,440)
\bezier{201}(340,440)(360,260)(340,260)
\bezier{244}(340,260)(320,220)(340,20)
\put(280,640){\vector(1,0){40}}
\put(440,640){\vector(-1,0){40}}
\put(240,620){$d$}
\put(396,399){$r_T\simeq \sqrt{\frac{\kappa\tau}{\rho C} }$}
\put(400,300){$N\simeq \frac{r_0^3}{r_T^3}$}
\end{picture}
\caption{Illustration for qualitative consideration of temperature 
thermodynamic fluctuations producing the
thermoelastic noise}\label{fig1}
\end{figure}
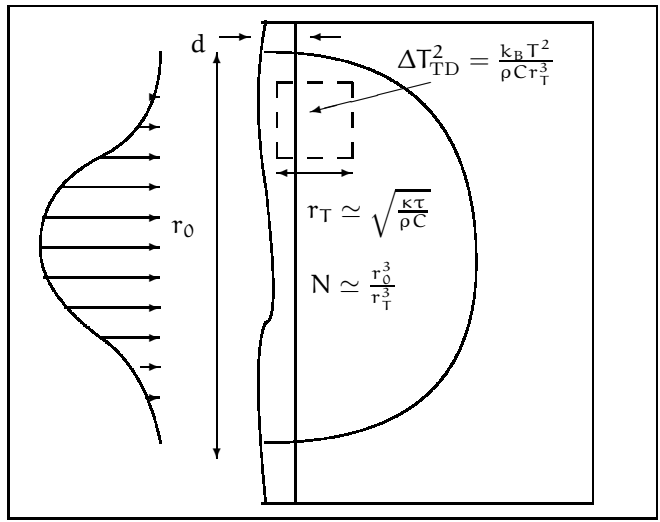

\section{Semi-qualitative consideration}\label{Ql}

\paragraph{"Bulk"  TD fluctuations.}
We have shown in \cite{bgv} that thermodynamical fluctuations of temperature 
in mirrors are transformed due to thermal expansion coefficient
$\alpha_b=(1/l)(dl/dT)$ into additional noise which can be a serious 
"barrier" limiting sensitivity. This noise (it is also called
as thermoelastic noise) produces random fluctuational "ripples" on surface. 
This noise has a nonlinear origin as nonzero value of
$\alpha_b$  is due to the lattice anharmonicity.

Our main interest here is fluctuations of 
averaged  displacement $\overline{X}(t)$ of mirror face surface. 
It is the normal
displacement $v_z$ of the mirror face averaged over the beam spot 
Gaussian power profile:
\begin{equation}\label{eq1}
\overline X = \frac{1}{\pi r_0^2}\int_0^\infty\int_0^{2\pi} e^{-r^2/r_0^2}
	v_z(r,\,\phi)\,r\,d\phi dr ,
\end{equation}
($r_0$ is the beam radius). The spectral density of displacement $\bar X$
can be presented for half infinite medium as follows:
\begin{equation}\label{eq2}
S_{\rm bulk}^{\rm TD}(\omega)=\frac{8}{\sqrt{2\pi}}
	\frac{\alpha^2_{b}k_{\rm B}T^2(1+\nu)^2\kappa}{
	 C_V^2r_0^3\omega^2}.
\end{equation}
 Here $k_{\rm B}$ is the Bolzmann constant, $T$ is temperature, $\nu$ is 
Poisson ratio, $\kappa$ is thermal conductivity  and $C_V$ is specific 
heat capacity per unit volume. This result has been refined for the 
case of finite sized mirror by Liu and Thorne \cite{liu}, however, the
 variation from our result is only several percent for typical mirror sizes, 
and hence we use more compact expression (\ref{eq2}) here.
 
 This physical result can be illustrated using the following qualitative 
consideration. We consider the surface  fluctuations
 averaged over the spot with radius $r_0$ which is larger than the 
characteristic diffusive heat transfer length $r_{\rm T}$
 \begin{equation}
 r_{\rm T}=\sqrt{\frac{\kappa}{C_V\omega}},\quad r_{\rm T}\ll r_0,
	\nonumber
 \end{equation}
where $\omega$ is the characteristic frequency (for LIGO it is about 
$\omega\simeq 2\pi\times100$~s$^{-1}$). Since in fused silica $r_{\rm
T}\simeq3.9\times10^{-3}$ cm and in sapphire 
$r_{\rm T}\simeq1.4\times10^{-2}$~cm the condition 
$r_{\rm T}\ll r_0\simeq 6$~cm is fulfilled (material
parameters are presented in Appendix \ref{A2}).

We know that in volume $\sim r^3_{\rm T}$ the 
length variation due to temperature TD 
fluctuations $\Delta T$ is about
\begin{equation}
 \Delta x_{\rm T}=\alpha_{b}\Delta T r_{\rm T}\simeq
	\alpha_br_{\rm T}
	\sqrt{\frac{k_{\rm B}T^2}{C_Vr^3_{\rm T}}}.\nonumber
\end{equation}
The TD temperature fluctuations in such volumes can be considered as 
independent ones. The number of such volumes that contribute to surface
fluctuations is about $N\simeq r_0^3/r_{\rm T}^3$, and hence the displacement 
$\overline X$ averaged over the spot with radius $r_0$
consists of these independently fluctuating volumes displacements sum, and is 
approximately equal to 
\begin{equation}
\label{eq3}
 \overline X\simeq\frac{\Delta x_{\rm T}}{\sqrt N}\simeq
	\alpha_{b}r_{\rm T}\sqrt{\frac{k_{\rm B}T^2}{C_V r^3_0}}.
 \end{equation}
Comparing (\ref{eq3}) with (\ref{eq2}) we can see that 
	$\overline X\simeq\sqrt{S^{\rm TD}_{bulk}\Delta\omega}$ 
(correct to the multiplier of about unity) if one assumes 
$\Delta \omega\simeq\omega$. This fact confirms that our semi-qualitative 
consideration is correct.

\paragraph{TD fluctuations in thin layer.}
One can consider the surface fluctuations of half space covered by thin layer, 
produced by TD temperature fluctuations as consisting of two
parts: "bulk" part depending on mirror body thermal expansion coefficient 
$\alpha_b$ (which has been already calculated in
\cite{bgv}) and "surface" part due to temperature fluctuations in layer with 
thickness $d$  and effective thermal expansion coefficient
$\alpha=\alpha_l-\alpha_b$ (here $\alpha_l$ is thermal expansion 
coefficient of layer  material).

It is important to note that layer thickness $d$ is much smaller than the 
diffusive heat transfer characteristic length $r_T$
(for optical coating $d\leq10\:\mu$m) and therefore the 
following condition is valid:
\begin{equation}
\label{cond}
r_0\gg r_T\gg d
\end{equation}
Therefore we may consider temperature 
fluctuations in our layer to be the same as in layer with
thickness $r_{\rm T}$. It means that TD temperature fluctuations in layer 
{\it does not} depend on thickness $d$. In this case we propose
the following semi-qualitative consideration:
\begin{eqnarray}
 \overline X_d &\simeq &\alpha\Delta Td,\\
 \Delta T &\simeq &\sqrt{\frac{k_{\rm B}T^2}{C_Vr^3_{\rm T}}}\times
	\sqrt{\frac{r^2_{\rm T}}{r^2_0}}=
	\sqrt{\frac{k_{\rm B}T^2}{C_Vr_0^2r_{\rm T}}},\label{eq4}\\
 \overline X_d &\simeq &\alpha d
	\sqrt{\frac{k_{\rm B}T^2}{C_Vr_0^2r_{\rm T}}}.
\end{eqnarray}
Formula (\ref{eq4}) is in good agreement with rigorous formula for spectral density $S_{\rm \Delta T}(\omega)$ of averaged temperature
$\Delta T$ in layer obtained in \cite{bgv2}
\begin{equation}\label{eq7}
S_{\rm \Delta T}(\omega)\simeq\frac{\sqrt2 k_{\rm B}T^2}{\pi r_0^2
	\sqrt{\rho C \kappa \omega}}=
	\frac{\sqrt2k_{\rm B}T^2}{\pi C_Vr_0^2r_{\rm T}\omega},
\end{equation}
and one can estimate the spectral density 
$S^{\rm TD\, est}_{\rm layer}(\omega)$ 
of surface fluctuations averaged over spot with radius $r_0$
as following
\begin{equation}\label{eq8}
S^{\rm TD\, est}_{\rm layer}(\omega)\simeq
	\alpha^2d^2S_{\rm \Delta T}(\omega)=\frac{\sqrt2\alpha^2d^2
	k_{\rm B}T^2}{\pi\rho Cr_0^2r_{\rm T}\omega}.
\end{equation}
Below we will show this formula to give the result correct to the multiplier 
of about unity.

\section{Analytical results}

Using Langevin approach \cite{bgv} (see also Appendix A) we have calculated 
the field of temperature TD fluctuations approximating mirror as a
half infinite medium with assumption that all material parameters of 
layer and half space are identical with exeption of thermal expantion
coefficients --- $\alpha_b$  for half space and $\alpha_l$ 
for layer.  These temperature fluctuations produce surface deformations 
of half infinite elastic space covered by thin layer. These
deformations can be calculated using elastic equation (\ref{fullandau}) 
(presented in Appendix A).

We use {\it quasistatic} approximation for elastic problem because time sound 
requires to travel across the distance $r_0$ is much smaller
than characteristic time $\simeq1/\omega$. We also assume  the layer to be 
homogeneous. We can divide our problem into two problems:

a) The "bulk" problem. The surface fluctuations are produced 
by TD temperature 
fluctuations in half space covered by layer with different
elastic constants. The layer and half space have the {\it same} thermal 
expansion coefficient $\alpha_b$. It corresponds to the
problem solved in \cite{bgv}.

b) The "surface" problem. The surface fluctuations are also produced by 
TD temperature fluctuations in half space covered by layer. The
effective layer thermal expansion coefficient 
$\alpha=\alpha_{\rm l}-\alpha_b\neq 0$, and there is 
{\it no thermal expansion in half space}.

The qualitative considerations in previous section make clear that TD 
temperature fluctuations producing fluctuations
in "surface" and "bulk" problems does not virtually  correlate and can 
be calculated independently.

For "surface" problem we assume that additional elasticity produced by layer 
is much smaller than mirror body bulk
elasticity (due to small thickness of the layer) and we take into account 
{\it only stresses} produced by layer due to
nonuniform temperature distribution $u(\vec r)$. Then one can take equation 
(\ref{fullandau}) and just substitute into right part the term
\begin{eqnarray}
\alpha\vec{\nabla}u &\Rightarrow& \alpha d\vec\nabla u\delta(x-\epsilon),\\
\alpha &=& \alpha_{l}-\alpha_{b}.
\end{eqnarray}
(Recall that here we assume {\it identity of material constants} for layer 
and half space.) As a result we obtain
spectral density of surface fluctuations (see details in Appendix \ref{A}):
\begin{equation}
\label{eq11}
S_{\rm layer}^{\rm TD}(\omega)=
	\frac{4\sqrt2(1+\nu)^2}{\pi}\frac{\alpha^2d^2k_{\rm B}T^2}{r_0^2
	\sqrt{\kappa C_V\omega}}.
\end{equation}
Here $\nu$ is Poisson ration.
Note that this formula differs from semi-qualitative one (\ref{eq8}) only by 
multiplier of about unity.

The fact that 
$S_{\rm layer}^{\rm TD}(\omega)>S_{\rm layer}^{\rm TD\: est.}(\omega)$ 
can be qualitatively explained by
the speculation that thermal expansion in layer also causes stresses in 
mirror body under layer: the material under
layer additionally "swells up" ("distend"). The similar effect one can 
observe when heated bimetallic plate is
bending.

We have generalized the formula (\ref{eq11}) for the case of {\em different 
material parameters} of layer and substrate deriving it from 
fluctuation-dissipation theorem (FDT) (see details in Appendix \ref{A2}): 
\begin{eqnarray}
\label{eq11a}
S_{\rm layer}^{\rm TD}(\omega)&=&
	\frac{4\sqrt2(1+\nu_b)^2}{\pi}
	\frac{\alpha^2d^2k_{\rm B}T^2}{r_0^2
	\sqrt{\kappa_b C_{V,b}\omega}},\\
\label{eq11b}
\alpha &=& \alpha_b\Lambda,\\
\label{eq11c}
\Lambda&=& - \frac{C_{V,\,l}}{C_{V,\, b}}+\frac{\alpha_l}{2\alpha_b} \times\\
&&\times	\left[\frac{1+\nu_l}{(1-\nu_l)(1+\nu_b)}+
	\frac{E_l(1-2\nu_b)}{E_b(1-\nu_l)}\right].\nonumber
\end{eqnarray}
Here $E_b,\ E_l$  are Young modulus, subscripts $_b$ and $_l$ refer 
to parameters of half space  and  layer correspondingly. This result was 
earlier obtained in \cite{marty}. 

We can further generalize the result (\ref{eq11a} - \ref{eq11c}) for 
multilayer coating with 
alternate material constants which differs from material constants of
mirror body. Let the multilayer
coating consists of $N$ alternating sequences of quarter-wavelength 
dielectric layers. Every odd layer has the
refraction index $n_1$, thickness $d_1=\lambda/4n_1$ ($\lambda$ is the 
incident light wavelength). Every even layer has parameters $n_2$,
$d_2=\lambda/4n_2$ correspondingly. Then formulas (\ref{eq11a}, \ref{eq11c}) 
can be used with the total coating thickness and effective expantion 
coefficient calculated using the following formulas (see details in 
Appendix \ref{A2}):
\begin{eqnarray}
\label{eq11d}
\alpha &=&
	\frac{\alpha_b\, d_{1}}{d_{1}+d_{2}}\,\Lambda_{1} +
	\frac{\alpha_b\, d_{2}}{d_{1}+d_{2}}\,\Lambda_{2}\\
\label{eq11e}
d &=& N(d_{1}+d_{2})
\end{eqnarray} 
Here factors $\Lambda_1$ and $\Lambda_2$ are calculated using (\ref{eq11c}) 
with substitution material parameters for odd  and even layers correspondingly.

It should be emphasized that negative influence of thermoelastic 
fluctuations in layer in {\it finite sized}
mirror can be larger than in half infinite space model due to 
"bimetallic" effect: the temperature fluctuations
in layer should additionally cause mirror bend through thermal expansion. 
Using FDT approach developed by Liu and Thorne \cite{liu} we have calculated
numerically  the following coefficient
\begin{equation}
C_{\rm fsm}=\frac{S_{\rm layer}^{\rm TD,\: finite\: test\: mass}}{
	S_{\rm layer}^{\rm TD,\: infinite\: test\: mass}}.
\end{equation}  
For design planned in LIGO-II \cite{ligo2} the test mass is manufactured 
from fused silica with $Ta_2O_5+SiO_2$  coating   
and has following parameters $R=19.4$ cm, $H=11.5$ cm, $r_0=6$ cm. 
For this case we estimate
$$
C_{\rm fsm}\simeq 1.56
$$ 
(see details of rather combersome calculations in Appendix \ref{A3}).
But if $R\gg H$ then the value of $C_{\rm FTM}$ may be substantialy larger.

\section{Numerical estimates}

High reflectivity of mirrors is provided by multilayer coating which 
consists of alternating sequences of
quarter-wavelength dielectric layers having refraction indices $n_1$ and 
$n_2$. The frequently used pairs are
$Ta_2O_5$ ($n_1\simeq 2.1$) and $SiO_2$ ($n_2\simeq 1.45$): namely these 
coatings are used in LIGO \cite{srini}. The
total typical for high finesse mirror  number of layer pairs is 
$N\simeq 19$ (this value we used for estimates below) so that total 
thickness of $Ta_2O_5+SiO_2$ coating is about $N(d_1+d_2)\simeq
6\times 10^{-4}$~cm. 

To estimate TD temperature fluctuations in interferometric coating of mirror
we use formula (\ref{eq11a}) with substitutions (\ref{eq11b} - \ref{eq11e}).
To obtain numerical estimates for these types of fluctuations it is necessary 
to have the value of $\alpha$ for thin
layer of $Ta_2O_5$. In the existing publications we have found very wide range:
from $\alpha_{\rm Ta_2O_5}\simeq-(4.43\pm 0.05)\times10^{-5}$ K$^{-1}$
\cite{inci} (the scheme of measurement is described in \cite{inci2}) to  
$\alpha_{\rm Ta_2O_5}\simeq 3.6\times10^{-6}$ K$^{-1}$ (without measurement 
error, unfortunately) \cite{tien}
and $\alpha_{\rm Ta_2O_5}\simeq (5\pm 2)\times10^{-6}$ K$^{-1}$ \cite{asam}. 
It can be explained not only by possible errors of experiment but also
by the fact that properties of tantalum pentoxide may be strongly depends on 
procedure of layer deposition on substrate \cite{stan}. 
So we use for estimate the
value $\alpha_{\rm Ta_2O_5}\simeq 5 \times10^{-6}$~K$^{-1}$  
because  coating measured in paper \cite{asam}  was  deposited on 
thin fused silica plates by the {\em same technology} as coating on 
LIGO mirrors. 
The other material parameters are listed in Appendix~\ref{A2}.

It is also useful to present estimates for thermorefractive noise \cite{bgv2}: 
TD temperature fluctuations produce
fluctuations of coating layers refraction indices $n_1$ and $n_2$ due to its 
dependence on temperature:
\begin{equation}
\beta_1=\frac{dn_1}{dT}\neq0,\quad \beta_2=\frac{dn_2}{dT}\neq0,\nonumber
\end{equation}
Refraction indices fluctuations produce in turn fluctuations of phase of the 
wave reflecting from mirror. This noise 
can be recalculated into equivalent noise displacement of 
mirror\footnote{Here we present formula correcting an error
in formula (3) in \cite{bgv2}}:
\begin{equation}\label{eq16}
S_{\rm trefr}^{\rm TD}(\omega)=\frac{\sqrt2\beta^2\lambda^2
	\kappa T^2}{\pi r_0^2\sqrt{\rho C\kappa\omega}},
\end{equation}
\begin{equation}\label{eq17}
\beta=\frac{n_1n_2(\beta_1+\beta_2)}{4(n_1^2-n_2^2)}.
\end{equation}
For estimates we use "optimistic" (i.e. smallest) 
value of $\beta_{\rm Ta_2O_5}\simeq 2.3 \times 10^{-6}$ K$^{-1}$ \cite{chu}
(for example, \cite{inci} reports unexpectedly high value 
$\beta_{\rm Ta_2O_5}\simeq 1.4 	\times 10^{-4}$ K$^{-1}$).

We also want to estimate the noise associated with the mirror material 
losses described in the model of structural 
damping \cite{saulson}, below we denote it as Brownian motion of the surface. 
In this model the angle of losses $\phi$ does
not depend on frequency and the following formula is valid for its spectral 
density in the infinite medium \cite{french,bgv,liu}:
\begin{equation}\label{eq18}
S_x^{\rm B}(\omega)\simeq\frac{4k_{\rm B}T}{\omega}\frac{(1-\nu^2)}{
	\sqrt{2\pi}Er_0}\phi.
\end{equation}
Note that we have used the new value of 
$\phi \simeq 0.5 \times 10^{-8}$ for the loss angle in very pure fused 
silica \cite{fsilica}.

The gravitational wave antenna spectral sensitivity to the perturbation 
of metric $h(\omega)$ may be recalculated from 
the displacement $X$ noise spectral density using the following formula:
\begin{equation}\label{eq19}
h(\omega)=\frac{\sqrt{N_0S_x(\omega)}}{L}
\end{equation}
where we use $N_0 = 4$ due to the fact that antenna has two arms 
(with length $L$) with two mirrors in each one. However, for 
TD layer fluctuations we use $N_0 \simeq 2$ because in LIGO interferometer 
arms the Fabry-Perot resonators end test 
mass (ETM) coating TD fluctuations are considerably larger than ones in the 
input test mass (ITM) coating due to the 
coating thickness in ETM is approximately $\sim 6$ times larger 
\cite{srini}.

\begin{figure}[t]
\psfrag{Silica}{\hspace{-2cm} Fused silica with $Ta_2O_5+SiO_2$}
\psfrag{SQL}{ SQL}
\psfrag{TD bulk}{TD bulk}
\psfrag{TD layer}{TD layer}
\psfrag{TD refr}{TD refr}
\psfrag{TD bulk}{TD bulk}
\psfrag{Brownian}{Brown}
\psfrag{frequency, Hz}{\hspace{10mm} frequency, Hz}
\psfrag{h, Hz^(1/2)}{\qquad  h, $1/\sqrt{Hz}$}
\psfrag{1e-25}[0.5]{$10^{-25}$}
\psfrag{1e-23}[0.5]{$10^{-23}$}
\psfrag{1e-24}[0.5]{$10^{-24}$}
\psfrag{10}[0.5]{10}
\psfrag{100}[0.5]{100}
\psfrag{1000}[0.5]{1000}
\includegraphics[width=0.5\textwidth]{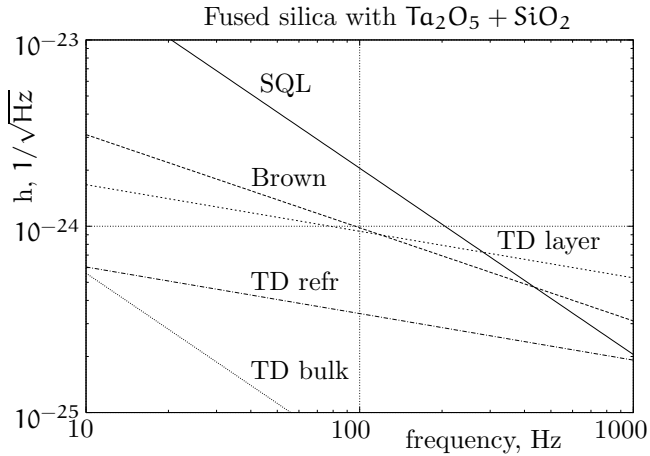}
\caption{Dependence of different noises on frequency in gravitational wave 
	antenna with fused silica mirror 
	and $Ta_2O_5 + SiO_2$ coating presented
	in terms of dimensionless metric (\ref{eq19}). Notations: "SQL" --- 
	$h_{\rm SQL}(\omega)$ (\ref{eq20}), "Brown" --- Brownian 
	fluctuations caused by structure losses and described by formula 
	(\ref{eq18}), "TD layer" - surface fluctuations 
	caused by TD temperature fluctuations (thermoelastic) 
	(\ref{eq11} -- \ref{eq11d}) in coating, "TD bulk" --- the 
	same fluctuations in the mirror volume (\ref{eq2}), "TD refr" --- 
	thermorefractive noise (\ref{eq16}).}\label{fig2}
\end{figure}

The  LIGO-II  antenna   will approach  the level of SQL so we also compare 
the noise limiting sensitivity with this 
limit in spectral form \cite{thorne}:
\begin{equation}
\label{eq20}
h_{\rm SQL}=\sqrt{\frac{8\hbar}{m\omega^2L^2}}.
\end{equation}
For estimate we use the set of parameters for fused silica mirror 
with $Ta_2O_5 + SiO_2$ coating using the parameters
listed in Appendix~\ref{param}. The results are presented in fig.~\ref{fig2}.

We can see that in fused silica mirror the TD bulk fluctuations are 
negligibly small but TD fluctuations in layer of
interferometric coating are about Brownian one and only 2 times smaller than 
standard quantum limit at $\omega
\simeq 2\pi \times 100$ s$^{-1}$.  Note that 
thermoelastic fluctuations in layer weakly depends on frequency 
$\sim \omega^{-1/4}$ (see (\ref{eq11})) and for  
frequencies $\ge  2\pi \times 300$ s$^{-1}$
these fluctuations dominate completely.

\begin{figure}[t]
\psfrag{Sapphire}{\hspace{-2cm} Sapphire with $Ta_2O_5+SiO_2$}
\psfrag{SQL}{\hspace{-3mm} SQL}
\psfrag{TD bulk}{TD bulk}
\psfrag{TD layer}{TD layer}
\psfrag{TD refr}{TD refr}
\psfrag{TD bulk}{TD bulk}
\psfrag{Brownian}{Brown}
\psfrag{frequency, Hz}{\qquad \quad frequency, Hz}
\psfrag{h, Hz^(1/2)}{\qquad h, $1/\sqrt{Hz}$}
\psfrag{1e-25}[0.5]{$10^{-25}$}
\psfrag{1e-24}[0.5]{$10^{-24}$}
\psfrag{1e-23}[0.5]{$10^{-23}$}
\psfrag{10}[0.5]{10}
\psfrag{100}[0.5]{100}
\psfrag{1000}[0.5]{1000}
\includegraphics[width=0.5\textwidth]{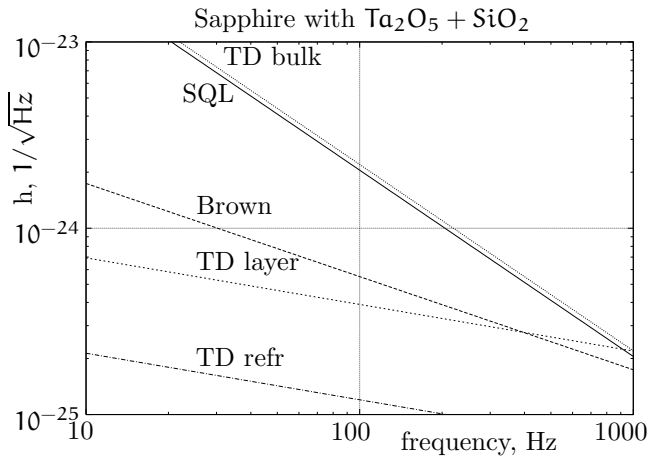}
\caption{Dependence of different noises on frequency in gravitational wave 
	antenna with sapphire mirror 
	and $Ta_2O_5 + SiO_2$ coating in terms of
	dimensionless metric (\ref{eq19}). Notations are the same as in 
	fig.\ref{fig2}}\label{fig3}
\end{figure}

For comparison we present in fig.~\ref{fig3} the estimates for sapphire 
mirror with the same $Ta_2O_5 + SiO_2$ 
coating. We can see that for sapphire mirror the contribution of TD noise 
in layer is smaller and only at $\omega
> 2\pi \times 10^3$ s$^{-1}$ it is close to SQL.

We can see that TD fluctuations in the same coating $Ta_2O_5 + SiO_2$ 
on sapphire and on fused silica   differs approximately by 3 times.
It can be explaied by the fact that factor $\kappa_s C_S$ (denominator
in (\ref{eq11})) for sapphire is about two orders larger than for 
fused silica. Note that equivalent thermal expansion coefficients described 
by formulas (\ref{eq11b} -- \ref{eq11d}) are practically the same
\begin{eqnarray*}
\alpha^{\rm Ta_2O_5 + SiO_2}_{\rm SiO_2}&\simeq & 2.8\times10^{-6}
	\:\rm K^{-1},\\
\alpha^{\rm Ta_2O_5 + SiO_2}_{\rm Al_2O_3}&\simeq & - 2.8\times10^{-6}
	\:\rm K^{-1},
\end{eqnarray*}

It is also worth noting that surface fluctuations caused by TD temperature 
fluctuations (thermoelastic and 
thermorefractive) in layer (coating)  depends on laser spot radius weaker
($\sim r_0^{-1}$) than "bulk" noise ($\sim r_0^{-3/2}$).

\section{Conclusion}
The slow dependence of interferometric layer (coating) TD noise on frequency 
$~\omega^{-1/4}$ and 
on the radius of the beam spot $~r_0^{-1}$ is worth noting.

We can see that TD noise in interferometric layer is close to the standard 
quantum limit and it may be an obstacle
for interferometric gravitational antennae (projects LIGO-II and especially 
LIGO-III). 

The resume of this paper may be formulated in the following way: it is 
important to measure {\em in situ} the  value of effective thermal 
expansion coefficient $\alpha$ for interferometric multilayer coatings
of high quality mirrors for presision measurements.

\section{Acknowledgements}
We are grateful for fruitful dicussions with M.~Fejer pointing out on 
incorrrectnes of several formulas in previous version of this paper.
We are pleased to thank Helena Armandula, Garilynn Billingsley, 
Sheila Rowan and Naci Inci providing priceless information
about material constants for optical coating. We are also grateful to  
F.~Ya.~Khalili and  V.~P.~Mitrofanov for helpful discussions.  

This work was supported in part by NSF and Caltech grant PHY0098715, by 
Russian Ministry of Industry and Science and
by Russian Foundation of Basic Researches.

\appendix

\section{TD fluctuations of surface of half space. Langevin approach}\label{A}

Here we present the derivation of formula (\ref{eq11}) using 
Langevin approach. We calculate the surface fluctuations (averaged over the 
spot with radius $r_0$) of half-space covered by
layer with thickness $d$. The surface fluctuations are originated by TD
temperature fluctuations which in turn produces deformations due to
non-zero thermal expantion. We consider the case when thermal expantion 
coefficient $\alpha$ is non-zero {\em only inside layer} and other thermal
and elastic parameters are the same in layer and half-space. 
We also assume that following strong unequalities  fulfills:
$$
d\ll r_T\ll r_0, \quad r_T=\sqrt{ \frac{\kappa}{\rho C \omega} }
$$

\subsection{TD temperature fluctuations}\label{sec:3}
 
We use Langevin approach proposed by \cite{bgv} using thermoconductivity 
equation for half-space  $z>0$ with thermoisolated surface 
with fluctuational forces $F$ in right part: 
\begin{eqnarray}
\label{TC}
\frac {\partial u}{\partial t}-a^2 \Delta u&=& F(\vec r,t),
	\quad a^2=\frac{\kappa}{\rho C},
	\\
\langle F(\vec r,t)F^*(\vec r_1,t_1)\rangle&=&
	 2\,\frac{k_B T^2 \kappa}{ (\rho C)^2 }\,	\delta(t-t_1)
	\times	\\ 
\times 
	\Delta \Big(\delta (x- x_1)\,\delta (y- y_1)&\times &	
	\big[\delta (z- z_1)+\delta (z+ z_1)\big]\Big).\nonumber 
\end{eqnarray}
Here we introduce mirror fluctuational forces (last term in square brackets)
in order to reformulate problem for the whole space. 

After space and time Fourier transform we can write down the 
solution
\begin{eqnarray} \label{u}
u(\vec r, t)&=&\int^{\infty}_{-\infty}\frac{d\vec k\, d\omega}{(2\pi)^4}\ u(\omega, \vec k)
	e^{i\omega t +i\vec k\vec r)},\label{ur}\nonumber\\
u(\omega, \vec k)&=&\frac{F(\vec k,\omega)}{a^2(\vec k )^2+ i\omega}
	\label{uk},\\
\label{corrFk}
\langle F(\vec k,\omega)F^*(\vec k_1,\omega_1)\rangle&=&
	 F_0^2T_0^2\,
	(2\pi)^4 |\vec k|^2\,
	\delta(\omega-\omega_1)\times \nonumber\\
&&\times \delta (k_x-k_{x1})\,\delta (k_y-k_{y1})\times \nonumber\\ 
&&\times \big[\,\delta (k_z-k_{z1})+\delta (k_z+k_{z1})\,\big].\nonumber\\
F_0^2T_0^2 &=& 2\,\frac{k_B T^2 \kappa}{ (\rho C)^2 }
\end{eqnarray}

\subsection{Equation of elasticity}

The problem of elastic deformations $\vec v$ for half-space covered
by thin layer caused by thermal expansion due to TD temperature fluctuations
$u$ can be described by equation
\cite{landau,boley} with zero stresses
$\sigma_{zz},\ \sigma_{yz},\ \sigma_{xz}$ on free surface $z=0$:
\begin{eqnarray}
\label {fullandau}
\frac{1-\nu}{1+\nu}&\times & \mbox{grad div } \vec v -\\
&& - \frac{1-2\nu}{2(1+\nu)} \mbox{rot rot } \vec v =
		\alpha \vec\nabla u,\nonumber\\
\sigma_{zz}&=&\frac{E}{1-2\nu}\left[
	\frac{\nu}{1+\nu}\left(
		\frac{\partial v_x}{\partial x}+
		\frac{\partial v_y}{\partial y}+
		\frac{\partial v_z}{\partial z}\right)\right. - \nonumber\\
&&\quad \left.\left.- \alpha u +
	\frac{1-2\nu}{1+\nu}\frac{\partial v_z}{\partial z} \right]
	\right|_{z=0}=0,\\
\sigma_{yz}&=&\left.\frac{E}{2(1+\nu)}\left(
		\frac{\partial v_z}{\partial y}+
		\frac{\partial v_y}{\partial z}\right)
	\right|_{z=0}=0, \\
\sigma_{xz}&=&\left.\frac{E}{2(1+\nu)}\left(
		\frac{\partial v_x}{\partial z}+
		\frac{\partial v_z}{\partial x}\right)\right|_{z=0}=0.
	\nonumber
\end{eqnarray}
Using condition $d\ll r_0$  we can present the right part of 
(\ref{fullandau}) as following 
\begin{eqnarray}
\label{alpha2}
\alpha  \vec\nabla u & \Rightarrow & \alpha d\,\vec \nabla \big(
	 u\, \delta(z-\epsilon)\big)
\end{eqnarray} 
One can find solution as sum (see for example sec. 3.4 in \cite{boley}):
$\vec v= \vec v^{(a)} +\vec \nabla \varphi $.
Here function $ \varphi $ fulfills Poison equation 
(it is derived from (\ref{fullandau})):
\begin{eqnarray}  
\label{varphiPua}
\Delta \varphi =\frac{1+\nu}{1-\nu} \alpha\, d\, u \delta(z-\epsilon)
\end{eqnarray}
without boundary condition and function $\vec v^{(a)}$ fulfills to
(\ref{fullandau}) with zero right part and the following boundary conditions:
\begin{eqnarray}
\sigma_{zz}&=&\left. \frac{E}{1+\nu}\left[
		\frac{\partial^2 \varphi}{\partial y^2}+
		\frac{\partial^2 \varphi}{\partial x^2} \right]
	\right|_{z=0},\label {boundaryva}\\
\sigma_{xz}&=&-\left.\frac{E}{(1+\nu)}\,
		\frac{\partial^2 \varphi}{\partial x \partial z}
		\right|_{z=0}, 	\nonumber\\
\sigma_{yz}&=&-\left.\frac{E}{(1+\nu)}\,
		\frac{\partial^2 \varphi}{\partial y \partial z}
        \right|_{z=0}. 	\nonumber
\end{eqnarray}
By this way for function $v^{(a)}$ we obtain the problem of half space 
deformation with stresses (\ref{boundaryva}) on boundary.

We are interesting of surface displacement along  $z$ axis averaged over 
spot with radius $r_0=\sqrt 2/\sigma_0$ in center of coordinate system:
\begin{eqnarray}
\label{barX}
\bar X(t)&=& \frac{1}{\pi\, r_0^2 } \int_{-\infty}^{\infty}dy\, dz\,
	\left.\left(\frac{\partial \varphi }{\partial z} +v_z^{(a)}\right)
	\right|_{z=0}\times\\
&&\qquad \times 
	\exp\left(-\frac{x^2+y^2 }{r_0^2}\right).\nonumber
\end{eqnarray}
Solution of Poison equation (\ref{varphiPua}) is known:
\begin{eqnarray}
\label{varphi}
\varphi (x,y,z) &=&-\frac{\alpha (1+\nu)}{4\pi (1-\nu)}
	\int \int_{-\infty}^{\infty}dx'dy'\times\\
&&\qquad \times
	\frac{2u(x',y',z')}{\sqrt{(x-x')^2 + (y-y')^2 +z^2}}.\nonumber
\end{eqnarray}
Here we add solution by mirror source ($\sim \delta(x +\epsilon)$) ---
it produces the multiplier $2$ in last fraction. 
Such function $\varphi$ makes zero contribution into surface displacement
(see (\ref{barX}). As addition the tangent stresses 
$\sigma_{xz}$ and $\sigma_{yz}$ in boundary conditions (\ref{boundaryva}) 
are also become zero. Only normal pressure is non-zero and is equal to:
\begin{eqnarray}
\sigma_{zz}^{z=0}&=&
	- \frac{\alpha d E(\partial_{yy}+\partial_{xx})}{2\pi (1-\nu)}
	\times \nonumber\\
&&\times 
         \int\int\int \int_{-\infty}^{\infty}
	\frac{dk_x dk_y dk_z d\omega}{(2\pi)^4}\times \nonumber\\
&&\quad \times 
        \frac{F(\vec k) T(\omega)}{a^2|\vec k|^2+ i\omega}\,
	e^{i\omega t+ik_y y+ik_x x} \times\nonumber\\
&&\qquad \times \int \int_{-\infty}^{\infty}dy'dx' 
	\times \nonumber\\
&&\qquad \quad \times 
	\frac {e^{ik_y (y-y')+ik_x (x'-x)}}
                {\sqrt{ (y-y')^2 +(x-x')^2}}\nonumber
\end{eqnarray}
We take the last integral (over $dy'dx'$)  going into 
cylindric coordinate system with notation $k_{\bot}=\sqrt{k_y^2+k_x^2}$
(using formulas 2.5.24.1 in \cite{int} and 2.12.4.28 in \cite{int1}):
\begin{eqnarray*}
I_1(k\bot) &=&\int \int_{-\infty}^{\infty}dy'dx' 
	\frac {e^{ik_y (y-y')+ik_x (x'-x)}}
                {\sqrt{ (y-y')^2 +(x-x')^2}}=\nonumber\\
&=&  \int_0^{2\pi}d\phi \int_{0}^{\infty}rdr \,
	\frac {e^{ik_{\bot}r\cos \phi}}
                {\sqrt{ r^2}}= \frac{2\pi}{k_\bot}
\end{eqnarray*}

Gathering we obtain expression for $\sigma_{zz}$
\begin{eqnarray}
\sigma_{zz}^{z=0}(y,x)&=& \frac{\alpha dE}{ (1-\nu)}\,
        \int\int\int \int_{-\infty}^{\infty}
	\frac{dk_x dk_y dk_z d\omega}{(2\pi)^4}\times\nonumber\\
&&\quad \times
        \frac{ k_\bot F(\vec k) T(\omega)}{a^2|\vec k|^2+ i\omega}\,
	e^{i\omega t+ik_y y+ik_x x} 
 \end{eqnarray}
Now we can write down the expression for $\vec v^{(a)}$ at $z=0$ 
(see sec. 8 in \cite{landau}):
$$
v^{(a)}_{z} = \frac{1-\nu^2}{\pi\, E}\int_{-\infty}^{\infty}dy'dx'
\frac{\sigma_{zz}^{z=0}(y',x')}{\sqrt{(y-y')^2+(x-x')^2}}.
$$
Substituting it into (\ref{barX}) we obtain:
\begin{eqnarray}
\label{barX(t)}
\bar X(t)&=& 
  2\alpha d (1+\nu)\,
        \int_{-\infty}^{\infty} \int_{0}^{\infty}
	\frac{dk_x\,dk_y\, dk_z\,  d\omega}{(2\pi)^4}\,\times \\
&&\times
        \frac{F(\vec k) T(\omega)}{(a^2|\vec k|^2+ i\omega)}
	\,	e^{i\omega t-\frac {k_{\bot}^2r_0^2}{4}}
	\nonumber
\end{eqnarray}
One can write down the time correlation function
(using (\ref{barX})):
\begin{eqnarray}
B(\tau) &=& \langle\, \bar X(t) \bar X(t+\tau) \,\rangle=\nonumber\\
&=&	4\alpha^2 d^2(1+\nu)^2\,
	\int_{-\infty}^{\infty} \frac{d\omega}{2\pi}\,
	e^{i\omega \tau}\times\\
&&\times         \int_{-\infty}^{\infty}\int_0^{\infty}
	\frac{dk_z \, k_\bot dk_\bot }{(2\pi)^2}\,
        \frac{2F_0^2 T_0^2k^2}{a^4|\vec k|^4+ \omega^2}\,
	e^{-k_\bot^2r_0^2/2}\nonumber
\end{eqnarray}
Now we can write down one side spectral density:
\begin{eqnarray}
\label{SX}
S_X(\omega)&=&  \frac{4}{\pi^2}\, (1+\nu)^2\alpha^2d^2\,F_0^2 T_0^2
	\times\\
&&\times     \int_{-\infty}^{\infty}\int_0^{\infty}
	\frac{(k_\bot^2+k_z^2)\,dk_z \, k_\bot dk_\bot }{(a^4 (k_\bot^2+k_z^2)^2+ \omega^2)}\,
	e^{-k_\bot^2r_0^2/2}\nonumber
\end{eqnarray}
Using condition $a^2/r_0^2 \ll \omega$ (and assuming that 
$k^2\simeq k_z^2$) we can easy take integral  over $k_\bot$ 
then over $k_z$ and obtain the formula (\ref{eq11}).

Note that formula (\ref{SX}) can be used for calculation $S_X$ with 
arbitrary  value $\omega$:
\begin{eqnarray}
\label{SX2}
S_X(\omega)&=&  \frac{4}{\pi^2}\, (1+\nu)^2\alpha^2d^2\,F_0^2 T_0^2
	\times I_X\\
I_X &=& \frac{1}{\sqrt 2 \, r_0 a^4}  \int_{-\infty}^{\infty}\int_0^{\infty}
	\frac{(x^2+y)\,dx \, dy }{(x^2+y)^2+ b^4)}\,
	e^{-y}=\nonumber\\
&=& \frac{\pi}{ 2 \, r_0 a^4}  \int_0^{\infty}
	\sqrt{\frac{y+\sqrt{y^2+b^4}}{(y^2+b^4)} }\ e^{-y}\, dy,\\
b^4 &=& \frac{\omega ^2 r_0^2}{4a^4}\nonumber,
\end{eqnarray}

\section{FDT approach  calculations of thermo-elastic noise\\
	 in interferometric coating}\label{A2}

Here we derive formula (\ref{eq11}) using fluctuation-dissipation theorem
(FDT)\cite{callenwelton,levin}
by the following thought experiment:

We imagine applying a sinusoidally
oscillating pressure,
\begin{equation}
P = F_o {e^{-r^2/r_o^2}\over \pi r_o^2 } \, e^{i\omega t}
\label{pressure}
\end{equation}
to one face half-infinite mass covered by layer. 
Here $F_o$ is a constant force amplitude, 
$\omega$ is the angular
frequency at which one wants to know the spectral
density of thermal noise, and  
the pressure distribution (\ref{pressure}) 
has precisely the same spatial profile as that of the generalized coordinate 
$\bar X$, whose thermal noise $S_{\bar X}(\omega)$ one wishes to compute.

The oscillating pressure $P$ feeds energy into the test mass, where it
gets dissipated by thermoelastic heat flow.
Computing the rate 
of this energy dissipation, $W_{\rm diss}$, averaged
over the period $2\pi/\omega$ of the pressure 
oscillations we can just write down (in according with fluctuation-dissipation
theorem) 
the spectral density of the noise $S_{\bar X}(\omega)$ is given by
\begin{equation}
S_{\bar X}(\omega) = {8 k_{\rm B}T W_{\rm diss} \over F_o^2 \omega^2}\;
\label{SqFD}
\end{equation}

The rate $W_{\rm diss}$ of thermoelastic dissipation is given by the
following standard expression (first term of Eq.\ (35.1) in \cite{landau}):
\begin{equation}
W_{\rm diss} =  \left\langle
\int {\kappa \over T} (\vec \nabla u)^2
\; r\, d\phi\, dr\, dz\right\rangle\;.
\label{wdissa}
\end{equation}
Here the integral is over the entire test-mass interior using cylindric
coordinates;
$T$ is the unperturbed temperature of the test-mass material and $u$ 
is the temperature perturbation produced by the oscillating pressure, 
$\kappa$ is the material's 
coefficient of thermal conductivity, and $\langle \dots \rangle$ denotes an
average over the pressure's oscillation period $ 2\pi/\omega$ 
(in practice it gives just a simple factor 
$\langle (\Re\,e^{i\omega t})^2 \rangle = 1/2$).

The computation of the oscillating temperature perturbation is made
fairly simple by two well-justified approximations \cite{bgv}: 

{\it First:} Quasistatic approximation. 
We can approximate
the oscillations of stress and strain in the test mass, induced by 
the oscillating pressure $P$, as {\it quasistatic}.  

{\it Second:} We assume that the following strong unequality
is valid 
$$
d\ll r_T \ll r_o,\quad r_T =\sqrt{ \frac{\kappa}{ \rho C\omega} }, 
$$ 
 here $r_T$ is  length of diffusive heat transfer and  $\rho$ is the density 
and $ C$ is the  specific heat . This 
means that, when computing the oscillating temperature distribution, we can  
approximate the oscillations of temperature as {\em
adiabatic} in transversal direction (tangent to surface)
but  {\em non-addiabatic} in normal to surface direction. 

The {\it quasistatic} approximation permits
us, at any moment of time $t$, to compute the test mass's internal displacement
field $\vec v$, and most importantly its expansion 
\begin{equation}
 \Theta={\rm div}\ \vec v,
\label{Theta}
\end{equation}
from the equations of static stress balance 
(Eq. (7.4) in \cite{landau}) 
\begin{equation}
(1-2\nu) \nabla^2 \vec v + \vec\nabla (\vec\nabla\cdot \vec v) = 0\;
\label{stressbalance}
\end{equation}
(where $\nu$ is the Poisson ratio), 
with the boundary condition that the normal pressure
on the test-mass face be $P(r,t)$ [Eq.\ (\ref{pressure})] and that all other
non-tangential stresses vanish at the test-mass surface. 

Ones $\Theta$ has
been computed, we can evaluate the temperature perturbation $u$ from
the thermoconductivity equation
\begin{equation}
\label{Theq1}
\Big(\partial _t-a^2\Delta\Big)u = 
	{- \alpha E T\over \rho C(1-2\nu)} 
	\partial _t\Theta,\quad a^2=\frac{\kappa}{\rho C} 
\end{equation}
Here $E$ is Young's modulus. It worth to underline that
we can not use addiabatic approach in (\ref{Theq1}) 
due to small thickness of layer. 

This temperature perturbation $u$ can then be plugged into Eq.
(\ref{wdissa}) to obtain the dissipation $W_{\rm diss}$ as an integral over
the gradient of the expansion.
This $W_{\rm diss}$ can be inserted into Eq.\ (\ref{SqFD}) to obtain the
spectral density of thermoelastic noise.

\subsection{Elastic problem}

\paragraph{Elastic infinite half space.}
We can assume that layer does not influece on deformations in substrate
due to its small thickness.
Then we can use the solution to the quasistatic stress-balance equation
(\ref{stressbalance})  given by a Green's-function
expression (see (8.18) in \cite{landau}) with $F_x = F_y = 0$, $F_z = P(r)$,
integrated over the surface of the test mass:
Below we will need the expression for $\Theta^{(b)}=\mbox{\bf div}\,\vec u $. 
Using results of \cite{bgv} one can calculate longitudial and
transversal parts of $\Theta^{(b)}$ separately:
\begin{eqnarray}
\label{Theta2} 
\Theta^{(b)}_\| &=& \left(\partial_x u_x +\partial_y u_y\right)_{z=0}=\\
&=& - \frac{2(1+\nu_b )(1-2\nu_b )P}{ E_b }\,
	,\\
\Theta^{(b)}_\bot &=& \partial_z u_z|_{z=0}=
	- \frac{(1+\nu_b )(1-2\nu_b )\, P }{ E_b }\, 
\end{eqnarray}

\paragraph{Layer.}
For solution of thermal problem we will need to calculate expansion 
$\Theta^{(l)}$ in layer. 
We assume that deformations of layer in transversal plane 
are the same as in substrate (the same natural assumption was done 
in \cite{marty}),
i.e. $\Theta^{(l)}_\|= \Theta^{(b)}_\||_{z=0}$.  One can use equation
(5.13) for stress in \cite{landau} for calculation 
$\Theta^{(l)}_\bot\equiv u_{zz}^{(l)}$:
\begin{eqnarray}
\label{sigmazz}
\sigma_{zz}&\equiv & - P=
	\frac{E_l }{(1+\nu_l )(1-2\nu_l )}\times\\
&&\times\Big((1-\nu_l )u_{zz}^{(f)}+
	\nu_l \underbrace{(u_{xx}^{(f)}+u_{yy}^{(f)})}_{\Theta^{(s)}_\||_{z=0}} 
	\Big).\nonumber
\end{eqnarray}
Using this equation one can find $u_{zz}^{(l)}$ and full expansion
$\Theta_l$ in layer:
\begin{eqnarray}
u_{zz}^{(l)} &=& -\frac{P}{Y_l (1-\nu_l )}\left(1-\frac{\nu_l Y_l }{Y_b }\right),\\
\Theta_l &=&
	-\frac{P}{Y_l (1-\nu_l )}\left(1+ \frac{Y_l (1-2\nu_l )}{Y_b }\right),\\
Y_b &=& \frac{E_b }{(1+\nu_b )(1-2\nu_b )},\quad 
	Y_l = \frac{E_l }{(1+\nu_l )(1-2\nu_l )}\nonumber
\end{eqnarray}

\subsection{Thermal problem}

The expansion in layer and in substrate are the source of heat with power 
per unit volume (see for example sec. 31 in \cite{landau}) which are equal 
to $W_l $
inside layer and to $W_b $ in substrate
$$
W_l = \frac{- \alpha_l E_l T}{ (1-2\nu_l )} \, i\omega\,\Theta_l ,
	\qquad 
	W_b = \frac{- \alpha_b E_b T}{ (1-2\nu_b )} \,i\omega\,\Theta_b .
$$ 
It seems that we can divide our problem into two problems:

{\em First:} The "bulk" problem. In half space there is power source $W_b $. 
	It corresponds to the problem solved in \cite{bgv}.

{\em Second:} The "surface" problem. There is power source only inside layer.
Below we consider only second problem.
 
We have thermal conductivity equation in layer with constant
right part. Using  conditions (\ref{cond})
one can simplify  thermal conductivity equations to one dimension form 
as following 
\begin{eqnarray}
\label{Theqs}
\Big(1 - \frac{\kappa_b }{i\omega\, C_b }\partial_z^2 \Big)\delta T_b &=& 0
	, \\
\label{Theqf}
\Big(1 - \frac{\kappa_l }{i\omega\, C_l }\partial_z^2 \Big)\delta T_l &=& w,
\end{eqnarray}
\begin{eqnarray}
w &=& \frac {- \alpha_l E_l T\,\Theta_l }{ (1-2\nu_l )\, C_l }+
	\frac{ \alpha_b E_b T\,\Theta_b }{ (1-2\nu_b )\, C_b }= \nonumber\\
&=&	-\frac{2(1+\nu_b )\, T\,\alpha_b }{ C_b }\, P\, \Delta,\\
\Delta &=& -1 +
	\frac{\alpha_l }{\alpha_b } \,\frac{C_b }{C_l }\,
	\frac{1+\nu_l }{2(1-\nu_l )(1+\nu_b )}\times\\
&&\times \left[1+
	\frac{E_l (1-2\nu_b )(1+\nu_b )}{E_b (1+\nu_l )}\right]
	\nonumber
\end{eqnarray}

We have condition of thermal isolation  at $z=0$ and condition of
continuity of temperature and heat flux at $z=d$. So we can find the solution 
of system (\ref{Theqs}, \ref{Theqf}) as following
\begin{eqnarray}
\delta T_l &=& A -B\cosh \gamma_l z,\quad \gamma_l =
	\frac{1+i}{\sqrt 2}\,\sqrt\frac{\omega\, C_l }{\kappa_l },\nonumber\\
\delta T_b &=& C\, e^{-\gamma_b (z-d)},\quad \gamma_b =
	\frac{1+i}{\sqrt 2}\,\sqrt\frac{\omega\, C_b }{\kappa_b },\nonumber\\
C &=& A-B\cosh\gamma_l d ,\\ 
\kappa_b \,\gamma_b \, C &=&\kappa_l \,\gamma_l \,B\, \sinh\gamma_l d ,\\
A&=&w,\quad C= 
	\frac{wR\, \sinh\gamma_l d}{R\,
			\sinh\gamma_l d+\cosh\gamma_l d},\\
B &=& \frac{w}{R\,\sinh\gamma_l d+\cosh\gamma_l d},\\
R &=& \frac{\kappa_l \gamma_l }{\kappa_b \gamma_b }
\end{eqnarray}
Due to conditions (\ref{cond}) and hence $|\gamma_g d|\ll 1$
we can simplify espressions for temperature:
\begin{eqnarray}
\delta T_l & \simeq & w(1-\cosh\gamma_l z+R\gamma_l d\cosh\gamma_l z) ,\\
\delta T_b &\simeq & 	\frac{w\,\kappa_l \gamma_l ^2\,d}{\kappa_b \gamma_b }\, 
	e^{-\gamma_b (z-d)}.
\end{eqnarray}
We are interesting only in temperature distribution in infinite space
because volume of layer produces small contribution into total budget of 
dissipated energy:
\begin{eqnarray} 
\label{wdiss}
W_{\rm diss}
&=& 	(1+\nu_b )^2\, \alpha_b ^2\,\frac{C_l ^2}{C_b ^2}\,\Delta^2	
	\frac{F_0^2}{\pi r_0^2}\times\\
&&\qquad \times \frac{\omega^2 d^2\, T}{\sqrt 2 \sqrt{C_b \kappa_b \omega}} 
	\nonumber
\end{eqnarray}
Using this result and (\ref{SqFD}) one can write down  spectral density 
\begin{eqnarray}
\label{Sx}
S_X(\omega) &=&
	\frac{4\sqrt 2 (1+\nu_b )^2\, 
		k_B T^2}{\pi r_0^2\sqrt{C_b \,\kappa_b \,\omega}}\, 
		\alpha_{\rm eff}^2d^2,\\
\alpha_{\rm eff} &=& \alpha_b \,\Lambda,\qquad 
	\Lambda=\frac{C_l }{C_b }\,\Delta.
\end{eqnarray}

\subsection{Generalization for multilayer coating}

We can generalize this result for multilayer coating consisting 
of $N$ alternating sequences of quarter-wavelength 
dielectric layers. Every odd layer has the
refraction index $n_1$, thickness $d_{1}=\lambda/4n_1$ ($\lambda$ is the 
incident light wavelength), its parameters we mark by subscript $_1$.
Every even layer has parameters $n_2$,
$d_{2}=\lambda/4n_2$,  its parameters we mark by subscript $_2$. It can be 
easy shown that the equivalent thermal expansion coefficient 
$\alpha_{\rm eff}$ of 
total coating and the total thickness $d$ of coating are 
the following:
\begin{eqnarray}
\label{eq15}
\alpha_{\rm eff} &=&
	\frac{\alpha_{s}\, d_{1}}{d_{1}+d_{2}}\,\Lambda_{1} +
	\frac{\alpha_{s}\, d_{2}}{d_{1}+d_{2}}\,\Lambda_{2},\\
d &=&N(d_{1}+d_{2}),\\
\Lambda_1 &=&- \frac{C_{f1}}{C_b }+
	\frac{\alpha_{f1}}{2\alpha_b } \times\\
&&\times	\left[\frac{1+\nu_{f1}}{(1-\nu_{f1})(1+\nu_b )}+
	\frac{E_{f1}(1-2\nu_b )}{E_b (1-\nu_{f1})}\right],
	\nonumber\\
\Lambda_2 &=& - \frac{C_{f2}}{C_b }+
	\frac{\alpha_{f2}}{2\alpha_b } \times\\
&&\times\left[\frac{1+\nu_{f2}}{(1-\nu_{f2})(1+\nu_b )}+
	\frac{E_{f2}(1-2\nu_b )}{E_b (1-\nu_{f2})}\right].\nonumber
\end{eqnarray}
And the spectral density $S_X(\omega)$ 
is defined by substitution (\ref{eq15}) into 
formula (\ref{Sx}). We have checked it by solution of thermal problem 
and calculation $W_{\rm diss}$ for one pair of layers, for two and 
three pairs of layers.

\section{Finite sized mirror}\label{A3}

Our calculations for finite sized mirror are based on results of Liu and 
Thorne \cite{liu}. The order of calculation steps is the following.
First step: we calculate total expansion  $\Theta_s$ and its tranversal part  
$\Theta_\|^{(s)}$ in substrate. 
Second step: we substitute calculated value $\Theta_\|^{(s)}$ into 
(\ref{sigmazz}) 
and calculate $\Theta_\bot^{(f)}$ and then $\Theta_f$. Third step: we 
substitute the values $\Theta_s$ and $\Theta_f$ into thermal cunductivity
equations (\ref{Theqs} and \ref{Theqf}). 
Forth step: we calculate dissipated energy $W_{\rm diss}$. 

\paragraph{Calculations of $\Theta_s$.} We use the 
results and notations of \cite{liu}. The radius of mirror is $R$, its
height is $H$.
\begin{eqnarray}
\label{ThetaFinite}
\Theta_s  & = & 	F_0 e^{i\omega t}\left(
	-\frac{p_0(1-2\nu_s)}{E_s} +
	\frac{2(1-2\nu_s)c_0}{E_s}+\right.\\
&&\left. +
	\sum_m [k_m A_m+B_m'] \,J_0(k_m r)\right),\nonumber\\
\lambda &=& \frac{E\nu}{(1-2\nu)(1+\nu)}\;, \quad
	\mu = \frac{E}{ 2(1+\nu)},\\
p_0 &=&\frac{1}{\pi R^2},\quad 
	c_0 = \frac{6\,R^2}{ H^2} 
	\sum_{m=1}^{\infty} \frac{J_0(\zeta_m) p_m }{ \zeta_m^2},
\end{eqnarray}
Here $\zeta_m $  are the m-th root of equation  $J_1(x)=0$,  
$J_0(x),\ J_1(x)$ are Bessel functions of zero and first orders.
\begin{eqnarray*}
p_m &=& \frac{\exp(-k_m^2 r_0^2/4)}{\pi R^2 J_0^2(\zeta_m)}, \quad 
	k_m=\frac{\zeta_m}{R},\\
A_m &=& A_m|_{z=0}=\gamma_m  + \delta_m,  	\label{Am} \\
B_m' &=& d_zB_m(z)|_{z=0}=
	k_m\, \frac{\mu_s(\beta_m-\alpha_m)}{(\lambda_s +2\mu_s )} -\nonumber\\
	&&\quad -k_m (\gamma_m+\delta_m),\\
D_m &=&[k_mA_m+B_m']=
	k_m\, \frac{\mu_s(\beta_m-\alpha_m)}{(\lambda_s +2\mu_s )},\\
\label{alphamTodeltam}
Q_m &=& \exp(-2 k_m H) \\
\alpha_m &=& \frac{p_m (\lambda_s +2\mu_s )}{k_m \mu_s (\lambda_s +\mu_s )}\
	\frac{1-Q_m+2 k_m H Q_m}{(1-Q_m)^2-4k_m^2 H^2 Q_m} \\
\beta_m &=& \frac{p_m (\lambda_s +2\mu_s )Q_m}{k_m \mu_s (\lambda_s +\mu_s )}\
	\frac{1-Q_m+2 k_m H}{(1-Q_m)^2-4k_m^2 H^2 Q_m} \\
\gamma_m &=& 
	\frac{-p_m}{2 k_m \mu_s (\lambda_s +\mu_s )}\, \frac{q_m^+\,Q_m+
	\mu_s (1-Q_m)}{(1-Q_m)^2-4k_m^2 H^2 Q_m}\nonumber \\
\delta_m &=& \frac{-p_m Q_m}{2 k_m \mu_s (\lambda_s +\mu_s )}\,
	\frac{q_m^- -\mu_s (1-Q_m)}{(1-Q_m)^2-4k_m^2 H^2 Q_m},\\
q_m^\pm &=& 2k_m^2 H^2 (\lambda_s +\mu_s ) \pm 2\mu_s k_m H,\\
D_m &=&	- \frac{p_m }{(\lambda_s +\mu_s )}\,
	\frac{(1-Q_m)^2}{(1-Q_m)^2-4k_m^2 H^2 Q_m}
\end{eqnarray*}

\paragraph{Calculations of $\Theta_\|^{(s)}$.}

\begin{eqnarray}
\label{ThetaT}
\Theta_\|^{(s)}  & = &	F_0 e^{i\omega t}\left(
	\frac{2\nu_s p_0}{E_s} +
	\frac{2(1-\nu_s)c_0}{E_s}+\right.\\
&&\left. +
	\sum_m k_m A_m|_{z=0} \,J_0(k_m r)\right),\\
k_mA_m &=& k_m(\delta_m+\gamma_m)=-\frac{p_m(1+\nu_s)}{E_s }\times\\
&&\times\left( 
	\frac{(1-Q_m)^2(1-2\nu_s)+4Q_m k_m^2 H^2 
		}{(1-Q_m)^2-4k_m^2 H^2 Q_m}\right).\nonumber
\end{eqnarray}

\paragraph{Calculations of expansion $\Theta_f$ in layer.}
Using (\ref{sigmazz}) we obtain:
\begin{eqnarray}
u_{zz} &=& \frac{-P(1+\nu_f)(1-2\nu_f)}{E_f(1-\nu_f)} - 
	\Theta^{(s)}_\|\, \frac{ \nu_f}{ 1-\nu_f},\\
\Theta_f &=& u_{zz}+\Theta^{(s)}_\|=\nonumber\\
&=&	\frac{-P(1+\nu_f)(1-2\nu_f)}{E_f(1-\nu_f)} + 
	\Theta^{(s)}_\|\, \frac{1-2 \nu_f}{ 1-\nu_f}
\end{eqnarray}
It is useful to present pressure $P$ as series and rewrite $\Theta_f$ as
following;
\begin{eqnarray}
P &=& 
	F_0\, e^{i\omega t} \left(	p_0+
	\sum_{m=1}^\infty p_m\, J_0(k_m r) \right),\\
\Theta_f &=& \frac{1-2 \nu_f}{ 1-\nu_f}\, F_0\, e^{i\omega t}\left(
	p_0\,\left[\frac{2\nu_s}{E_s}- \frac{(1+\nu_f)}{E_f}\right]+
	\nonumber\right.\\
&&\left. +
	\frac{2(1-\nu_s)c_0}{E_s}-
	\sum_{m=1}^\infty p_mJ_0(k_m r)K_m \right),\\
K_m &=& \frac{(1+\nu_f)}{E_f}+\frac{(1+\nu_s)}{E_s}\times \\
&&\times\frac{\Big[(1-Q_m)^2(1-2\nu_s)+4Q_m k_m^2 H^2 
		\Big]}{(1-Q_m)^2-4k_m^2 H^2 Q_m}  \nonumber
\end{eqnarray}

\paragraph{Calculations of right part in thermal conductivity equation.} 
Using (\ref{Theqf}) we obtain the expresion for $w$:
\begin{eqnarray}
w &=& \frac{- \alpha_f E_f T\,\Theta_f}{ (1-2\nu_f)\, C_f}+
	\frac{ \alpha_s E_s T\,\Theta_s}{ (1-2\nu_s)\, C_s},\\
C_f w&=&	\alpha_s TF_0e^{i\omega t}\times\\
&\times&\left\{ -p_0\left(
	\frac{C_f}{C_s} +\frac{\alpha _f}{\alpha_s}\left[
		\frac{2\nu_sE_f}{(1-\nu_f)E_s} -\frac{1+\nu_f}{1-\nu_f}\right]
	\right)+\right.\nonumber\\
&&+
	2c_0\left(\frac{C_f}{C_s}-
		\frac{\alpha_f E_f(1-\nu_s)}{\alpha_s E_s(1-\nu_f)}\right)+
	\nonumber\\
&&+\left.	\sum_{m=1}^\infty p_mJ_0(k_mr)L_m\right\},\nonumber\\
L_m &=&  \frac{\alpha_f(1+\nu_f)}{\alpha_s (1-\nu_f)}-\frac{1+\nu_s}{1-\nu_f}+
	\nonumber\\
&&+	\frac{(1+\nu_s)(1-Q_m)^2}{(1-Q_m)^2-4k_m^2 H^2 Q_m}\nonumber\times\\
&&\quad \times
	\left(
	\frac{\alpha_fE_f(1-2\nu_s)}{\alpha_sE_s(1-\nu_f)}+
	\frac{1}{1-\nu_f} -\frac{2C_f}{C_s}
	\right)
\end{eqnarray}

\paragraph{Calculations of dissipated energy $W_{\rm diss}$.}
Using (\ref{wdissa}) we obtain:
\begin{eqnarray} 
W_{\rm diss}^{\rm fsm}
\label{Wftm}
&=& \frac{\alpha_s^2 F_0^2\,T \,\omega^2
	d^2}{2\sqrt 2 \, \pi R^2\sqrt{\kappa_s C_s\omega}}\, {\cal P},\\
{\cal P} &=& 
	\left[
	\frac{\alpha _f}{\alpha_s}\left(\frac{1+\nu_f}{1-\nu_f}
		-\frac{2\nu_sE_f}{(1-\nu_f)E_s} \right)
	-\frac{C_f}{C_s}+\right.\nonumber\\
&&\quad +\left. 	S_1\left(\frac{C_f}{C_s}-
		\frac{\alpha_fE_f(1-\nu_s)}{\alpha_s E_s(1-\nu_f)}
	\right)\right]^2+S_2,\nonumber\\
S_1	&=& \frac{12R^2}{H^2}
	\sum_{m=1}^\infty 
	\frac{e^{-k_m^2r_0^2/4}}{\zeta_m^2\, J_0(\zeta_m)},\\
S_2 &=& \sum_{m=1}^\infty 
	\frac{e^{-k_m^2r_0^2/2}}{ J_0^2(\zeta_m)}\, L_m^2 .
\end{eqnarray}
Here we used the (nonstandard) orthogonality relations:
\begin{eqnarray*}
\int_0^R J_0(k_mr)\, J_0(k_nr)\, r\, dr&=&\frac{R^2}{2}\, J_0^2(\zeta_m)\,
	\delta_{mn},\\
\int_0^R J_0(k_mr)\,  r\, dr &=& 0.
\end{eqnarray*}

Now one can calculate numerically the coefficient
\begin{eqnarray}
C_{\rm fsm}&=&\sqrt\frac{W_{\rm diss}^{\rm fsm}}{W_{\rm diss}}=
	\sqrt\frac{r_0^2 {\cal P}}{2 R^2(1+\nu_s)^2\Lambda^2}\,
\end{eqnarray}  
using (\ref{Wftm}) and (\ref{wdiss}).

We numerically 
estimate $C_{\rm ftm}$ for $R=19.4$~cm, $H=11.5$~cm, $r_0=6$~cm and for
$Ta_2O_5+SiO_2$ coating on fused silica substrate:
$$
C_{\rm ftm}\simeq 1.56
$$

\section{Parameters}\label{param}

\begin{eqnarray*}
\omega&=&2\pi\times 100\ \mbox{s}^{-1},\quad
	T=300\ \mbox{K}, \quad r_0\simeq 6\ {\rm cm} \nonumber\\
m&=&3\times 10^4\ \mbox{ g},\quad \lambda = 1.06 \ \mu ,\quad 
	L=4\times 10^5\ \mbox{cm}; \label{parameter}; \\
&&\mbox{\bf Fused silica:}\\
\quad \alpha&=&5.5\times  10^{-7}\ \mbox{K}^{-1}, \ \
	\kappa=1.4\times  10^5\ \frac{\mbox{erg}}{\mbox{cm s K}},
	\label{silica}\\
\rho&=&2.2\ \frac{\mbox{g}}{\mbox{cm}^3}, \ \
	C=6.7\times  10^6\ \frac{\mbox{erg}}{\mbox{g K}},
	\nonumber \\
E&=&7.2\times 10^{11}\frac{\mbox{erg}}{\mbox{cm}^3},  \ \
	\nu=0.17,\\
\phi&=& 0.5\times10^{-8}\ \cite{fsilica}; \nonumber \\
\frac{d E}{dT} &=& -1.5\times 10^{8}\,\frac{\mbox{erg}}{\mbox{K\, cm}^3}.\\
n&=& 1.45,\ \ \frac{dn}{dT} = 1.5\cdot 10^{-5}\ \mbox{K}^{-1},\\
&&\mbox{\bf Sapphire:}\\
\quad \alpha&=& 5.0\times  10^{-6}\ \mbox{K}^{-1}, \ \
	\kappa=4.0\times  10^6\ \frac{\mbox{erg}}{\mbox{cm s K}},
	\label{saphire}\\
\rho&=&4.0\ \frac{\mbox{g}}{\mbox{cm}^3},\ \
	C=7.9\times  10^6\ \frac{\mbox{erg}}{\mbox{g K}},
	\nonumber\\
E&=&4\times 10^{12}\,\frac{\mbox{erg}}{\mbox{cm}^3},\ \
	\nu=0.29,\ \
	\phi=3\times10^{-9},\\
\frac{d E}{dT} &=& -4\times 10^{8}\,\frac{\mbox{erg}}{\mbox{K\, cm}^3},\\
n&=& 1.76,\quad (\lambda=1\ \mu m\ \cite{kikoin}\\
&&\mbox{\bf Tantal pentoxide}\ Ta_5O_5\\
\alpha&=&-4.4\cdot 10^{-5}\  \mbox{K}^{-1}\ \cite{inci}, \\
\alpha&=&3.6\cdot 10^{-6}\  \mbox{K}^{-1}\ \cite{tien},\\
\alpha &=& (5\pm 1)\cdot 10^{-5}\  \mbox{K}^{-1}\ \cite{asam}\\ 
E&=&1.4\times 10^{12}\,\frac{\mbox{erg}}{\mbox{cm}^3}\,\cite{martin}, \ \
	\nu=0.23,\ \cite{martin}\\
\kappa &=& 4.6\times  10^{6}\ \frac{\mbox{erg}}{\mbox{cm\, s\. K}},\ \
	\rho = 6.85\ \frac{\mbox{g}}{\mbox{cm}^3}\ ,\\
C &=& 3.06\times 10^6\ \frac{\mbox{erg}}{\mbox{g K}},\\
n&=& 2.1,\\ 
\frac{dn}{dT} &=& 1.21\cdot 10^{-4}\ \mbox{K}^{-1}\
	(\lambda\simeq 1.4\mu m)\ \cite{inci}\\
\frac{dn}{dT} &=& 2.3\cdot 10^{-6}\ \mbox{K}^{-1}\
	(\lambda\simeq 0.63\mu m) \cite{chu},\\
\frac{dn}{dT} &=& 4.7\cdot 10^{-6}\ \mbox{K}^{-1}\
	 \cite{scobey}
\end{eqnarray*}


\begin{thebibliography}{99}

\bibitem{bgv} V.\ B.\ Braginsky, M.\ L.\ Gorodetsky, and S.\ P.\ Vyatchanin,
{\em Physics Letters} {\bf A 264}, 1 (1999); cond-mat/9912139;


\bibitem{rempe} G. Rempe et al, Opt.Letters, {\bf 17}, 363 (1992).  

\bibitem{kimble}	H. J. Kimble, private communication.

\bibitem{bh1} V. B. Braginsky and F. Ya. Khalili, {\em Quantum Measurement},
	ed. by K.S.~Thorne, Cambridge Univ. Press, 1992.

\bibitem{bh2} V. B. Braginsky and F. Ya. Khalili, Rev. Mod. Physics, 
	{\bf 68}, 1 (1996).

\bibitem{abr1} A.\ Abramovici {\it et al.}, Science {\bf 256}, 326 (1992).
\bibitem {abr2} A.\ Abramovici {\it et al.}, Phys.Letters.A {\bf 218}, 
	157 (1996).
\bibitem{bru} V. B. Braginsky, Physics ``Uspekhi'', {\bf 43}, 691 (2000). 

\bibitem{bgv2} V.\ B.\ Braginsky, M.\ L.\ Gorodetsky, and S.\ P.\ Vyatchanin,
	{\em Physics Letters } {\bf A 271}, 303-307 (2000).

\bibitem{marty} M. M. Fejer, S. Rowan et al tbd, "Thermoelastic dissipation in 
	inhomogenious media: loss measurement and displacement noise \dots",
	{\em manuscript}, 2003

\bibitem{ligo2} 
	{\sf http://www.ligo.caltech.edu/$\sim$ligo2/\\
	scripts/12refdes.htm}

\bibitem{liu} Yu. T. Liu and K. S. Thorne, {\em Phys. Rev.} {\bf D62}, 122002
	(2000).

\bibitem{inci} M. N. Inci, Simultaneous measurements of thermal optical 
	and linear thermal expansion coefficients of $Ta_2O_5$ films, 
	{\em ICO 19}, 25-30 August 2002, Firenze, Italy.

\bibitem{inci2} G. Gulsen and M. N. Inci, {\em Optical Materials} 
	{\bf 18}, 373 (2002).

\bibitem{tien} C.-L. Tien, C.-C. Jaing, C.-C. Lee and K.-P. Chuang,	
	{\em J. Mod. Opt.}, {\bf 47}, 1681 (2000).

\bibitem{asam} V.~B.~Braginsky and A.~A.~Samoilenko, submitted to 
	{\em Physics Letters A}.

\bibitem{stan} Stan Whitcomb, private communication.

\bibitem{chu} A.K. Chu, H.C. Lin, W.H. Cheng {\em J. Electro. Mater.},	
	{\bf 26}, 889 (1997).


\bibitem{saulson} P. R.\ Saulson, {\em Phys. Rev. D}, {\bf 42}, 2437 (1990);\\
	G. I.\ Gonzalez and P. R.\ Saulson, {\em J. Acoust. Soc. Am.},
	{\bf 96}, 207 (1994).

\bibitem{french} F. Bondu, P. Hello, Jean-Yves Vinet,
	{\em Physics Letters} {\bf A 246}, 227 (1998).

\bibitem{srini} K. Srinivasan, Coating strain induced distortion in LIGO
	optics, LIGO document T-970176-00, 
	available on {\sf www.ligo.caltech.edu}.

\bibitem{thorne} V. B.\ Braginsky, M. L.\ Gorodetsky, F. Ya.\ Khalili and
        K. S.\ Thorne, {\em Report at Third Amaldi Conference, Caltech}, July,
        1999.

\bibitem {landau} L. D. Landau and E. M. Lifshitz\\
	{\em Theory of Elasticity}, Nauka, Moscow, 1980,
	({\em translation:} Pergamon, Oxford, 1986)

\bibitem{fsilica} A. Ageev, S. Penn and P.Saulson, Private communication. 
	E-mail: ageev\verb'@'physics.syr.edu


\bibitem {boley} B. A. Boley and J. H. Weiner\\
	{\em Theory of thermal stresses,} J.Wiley and sons, 1960,

\bibitem{int} 
	A. P. Prudnikov, Yu. A. Brychkov and O. B. Marychev,
	{\em Integrals and Sums}. Moscow, Nauka, 1983.

\bibitem{int1} 
	A. P. Prudnikov, Yu. A. Brychkov and O. B. Marychev,
	{\em Integrals and Sums. Special Functions}. Moscow, Nauka, 1983.

\bibitem{callenwelton} H.\ B.\ Callen and T.\ A.\ Welton, {\em Phys. Rev.} 
	{\bf 83}, 34 (1951). 

\bibitem{levin} Yu.\ Levin, {\em Phys. Rev.}\ D {\bf 57}, 659 (1998). 

\bibitem{martin}  P.J. Martin et al {\em Thin solid films} {\bf 239} 
	181-185 (1994)

\bibitem{scobey} M.A. Scobey and P. Stupik, {\em 37-th Annual Thechnical 
	Conference Proceedings,} 47-52 (1994).

\bibitem{kikoin} Physical parameters, {\em ed.} I.S. Grigorev and E.Z. Meilikhova,
	Moscow, 1991 ({\em in russian})
\end{thebibliography}
\end{document}